\documentclass[aps,prl,comment,twocolumn,superscriptaddress,groupedaddress]{revtex4}  
\usepackage{graphicx}  
\usepackage{dcolumn}   
\usepackage{bm}        
\usepackage{amssymb}   
\usepackage{comment}   
\usepackage{subfigure}   
\usepackage{amsmath}

\begin{document}

\title{Comment on ``Tunable Supermode Dielectric Resonators for Axion Dark-Matter Haloscopes"}
\date{\today} 

\author{Jinsu Kim} \affiliation{Department of physics, KAIST, Daejeon 34141 Republic of Korea} 
\author{SungWoo Youn}\email[Corresponding author: ]{swyoun@ibs.re.kr}\affiliation{Center for Axion and Precision Physics Research, IBS, Daejeon 34051 Republic of Korea}
\author{Junu Jeong} \affiliation{Department of physics, KAIST, Daejeon 34141 Republic of Korea} 
\author{Yannis K. Semertzidis} \affiliation{Department of physics, KAIST, Daejeon 34141 Republic of Korea} \affiliation{Center for Axion and Precision Physics Research, IBS, Daejeon 34051 Republic of Korea}

\maketitle

We comment on a recently published paper~\cite{bib:report}, which presents frequency-tuning mechanisms for dielectric resonators and demonstrates their potential application to axion haloscopes.
One of the schemes introduces a cylindrical dielectric hollow and splits it in the axial direction to tune the frequency.
The authors claim that this scheme offers a 1 to 2-order-of-magnitude improvement in axion search sensitivity in exploiting a higher-order resonant mode.
We find that their study is based on unrealistic cavity modeling and inappropriate choice of the figure of merit (FOM), which could mislead to the significant improvement in sensitivity.
Considering a practical cavity structure and an appropriate FOM, we recalculate the significance of the scheme, which turns out to be not substantial.

It is noted that the cavity in their simulation study is not realistically modeled to accommodate the tuning mechanism that they pursue in the paper.
For the realistic cavity design, the proposed tuning mechanism inevitably requires circular hollow holes on both the top and bottom end-caps of the resonator, through which a significant field leakage can be induced for transverse-magnetic modes.
Due to this, the calculated improvement in axion sensitivity could be overestimated.
Furthermore, we also notice that a reasonable axion sensitivity is not estimated because of their choice of the FOM, where the geometry factor ($G$) was chosen rather than the quality factor ($Q$).
A geometry factor, defined in the paper as
\begin{equation*}
G=\frac{\omega\mu_0\int|\vec{H}|^2dV}{\int|\vec{H}|^2dS}\; {\rm with} \; Q_s=\frac{G}{R_s},
\label{eq:geo_fact}
\end{equation*}
describes surface loss and is consistent with the quality factor only for closed geometries.
A quality factor, on the other hand, reflects other effects, such as field leakage and external interference, in the form of 
\begin{equation*}
\frac{1}{Q_{\rm tot}}=\frac{1}{Q_{\rm surf}}+\frac{1}{Q_{\rm leak}}+\frac{1}{Q_{\rm ext}}+\cdot\cdot\cdot,
\end{equation*}
and works for generic geometries.
Hereby, we repeat the calculations using a qualtity-factor-based FOM, $C^2V^2Q$, which is appropriate for axion search sensitivity.

\begin{figure}[h]
\centering
\subfigure[]{
\includegraphics[width=0.4\textwidth]{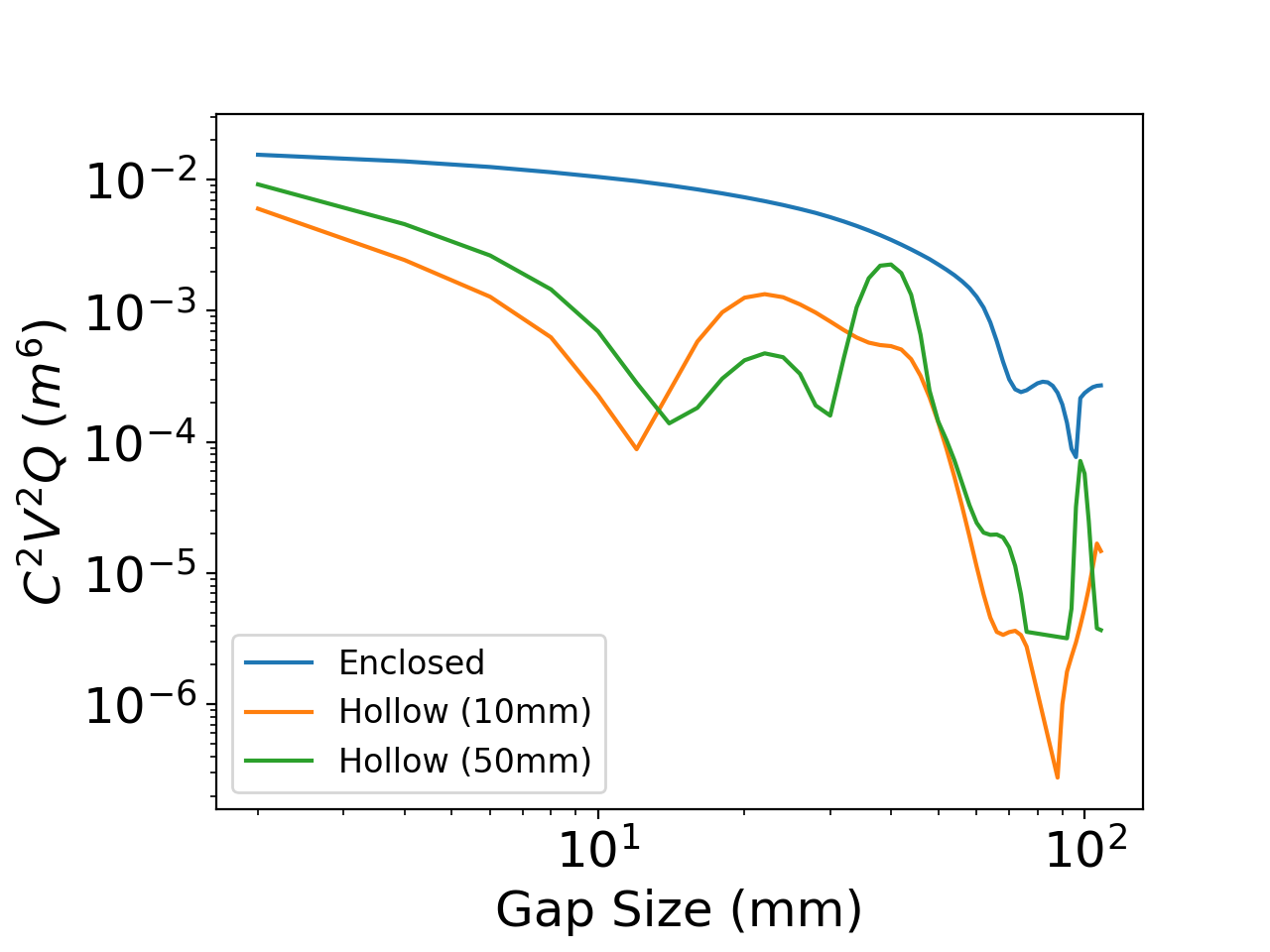}\label{fig:comp_fom}}
\subfigure[]{
\includegraphics[width=0.4\textwidth]{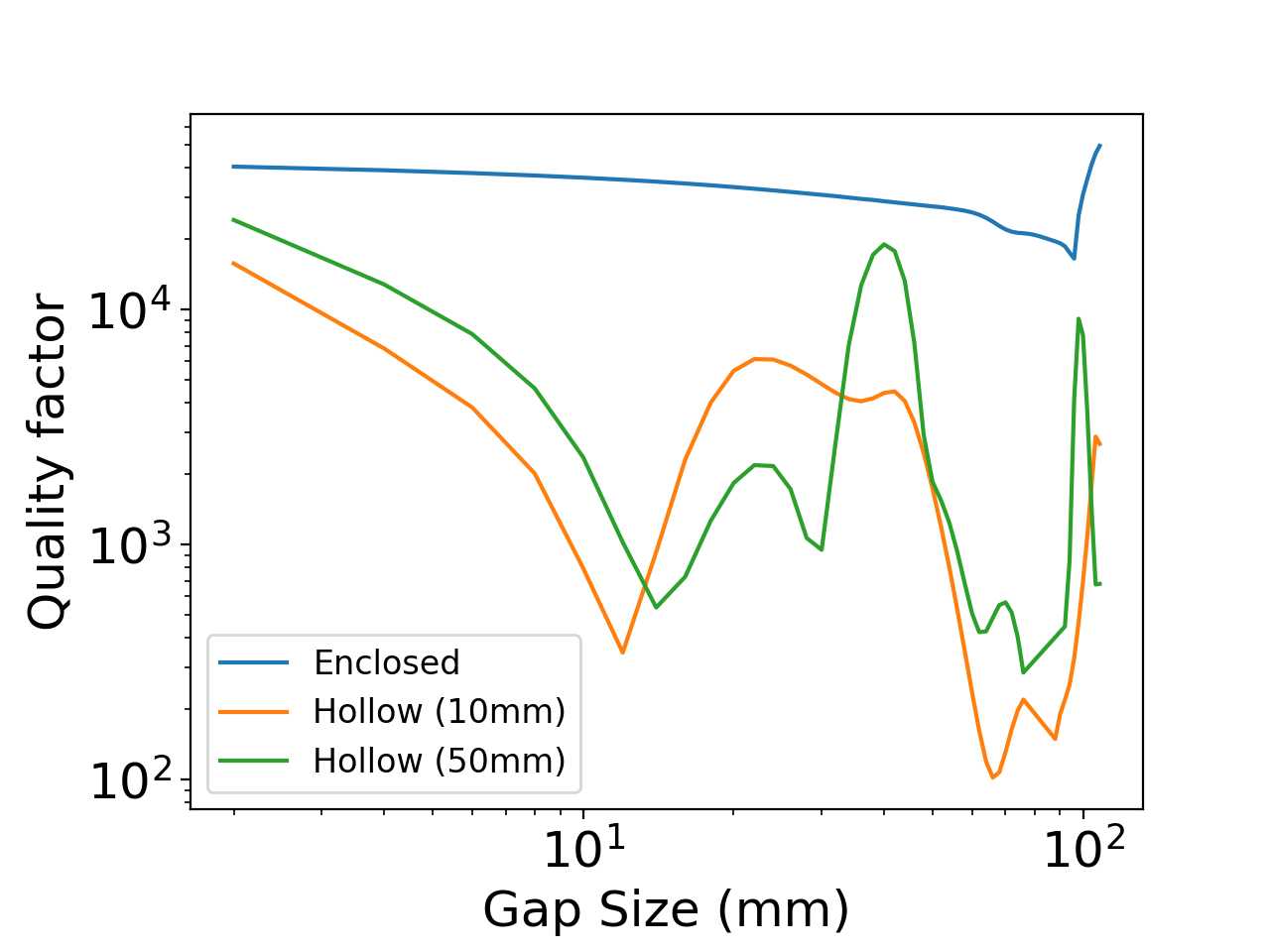}\label{fig:comp_q}}
\caption{Figure of merit ($C^2V^2Q$) (a) and quality factor ($Q$) (b) as a function of the gap size for various cavity designs.
The blue lines represent the enclosed (no hollow holes) design described in the original paper, while the orange and green lines represent the designs with 10\,mm and 50\,mm circular holes, respectively.}
\label{fig:comp}
\end{figure}

A simulation study is performed employing the same simulation software, cavity modeling, and dielectric material as those in the paper. 
Using the FOM based on the geometry factor, we obtain the consistent results with those described in Fig. 9 of the paper, which convinces that we understand their scheme. In the same configuration, we repeat the calculation using the FOM based on the quality factor. 
Now a realistic cavity design is considered by introducing a circular hollow gap on both the top and bottom end-caps of the resonator.
In order to see the effect of depth of the hollow gaps, two thicknesses of the end-caps are considered - 10\,mm and 50\,mm. 
Scattering boundary conditions are applied for non-metallic surfaces.
The results show that the realistic cavity design undergoes substantial reduction in FOM comparing to that for the unrealistic design described in the paper, as can be seen in Fig.~\ref{fig:comp_fom}, and that the improvement is not as significant as it was claimed.
It is seen that the main effect comes from a significant drop in quality factor, as shown in Fig.~\ref{fig:comp_q}.
There are also some indications that mode mixings may take place over the process of the frequency tuning.

In summary, we review the tuning mechanism for dielectric ring resonators proposed in Ref.~\cite{bib:report} and recalculate the sensitivity by considering a practical cavity design and an appropriate FOM.
The results reveal that the improvement in axion sensitivity is not as substantial as claimed in the paper.

This work was supported by IBS-R017-D1-2018-a00/IBS-R017-Y1-2018-a00.

\end{document}